\documentclass[trackchanges,twocolumn]{aastex701}

\begin{document}

\title{The Dyson Minds 2025 Workshop: SETI around Black Holes}

\author[0000-0002-0212-4563]{Olivia Curtis}
\affiliation{Department of Astronomy \& Astrophysics, The Pennsylvania State University, 251 Pollock Road, University
Park, PA 16802, USA} \affiliation{Penn State Extraterrestrial Intelligence Center, The Pennsylvania State University, 525 Davey Laboratory, 251 Pollock Road, Penn State, University Park, PA 16802, USA}
\email[show]{ocurtis@psu.edu} \affiliation{Institute for Gravitation and the Cosmos, The Pennsylvania State University, University Park, PA 16802}

\author[0000-0002-1812-4023]{Van Hunter Adams}
\affiliation{College of Electrical and Computer Engineering, Cornell University, 412 Computing and Information Sciences Building, Ithaca, NY 14853, USA}
\email{vha3@cornell.edu}

\author[0000-0001-6138-8633]{Daniel Angerhausen}
\affiliation{SETI Institute, 189 N. Bernado Ave, Mountain View, CA 94043, USA}
\affiliation{Blue Marble Space Institute of Science, Seattle, 600 1st Avenue, WA 98104, USA}
\affiliation{Institute for Particle Physics \& Astrophysics, ETH Zurich,  Wolfgang-Pauli-Str. 27, 8093 Zurich, Switzerland}
\email{daniel.angerhausen@gmail.com}

\author[]{Joseph Bates}
\affiliation{The Ultraintelligence Foundation}
\email{aih@aiforhumanityfoundation.org}

\author[0000-0003-2813-8469]{Anamaria Berea}
\affiliation{Department of Computational \& Data Sciences, George Mason University, 4400 University Drive, MS 6A12, Fairfax, Virginia, USA}
\email{aberea@gmu.edu}

\author[0009-0006-8928-3562]{Steven J. Dick}
\affiliation{Former NASA Chief Historian}
\email{stevedick1@comcast.net}

\author[0000-0001-5060-1398]{Martin Elvis}
\affiliation{Harvard-Smithsonian Center for Astrophysics, 60 Garden St., Cambridge, MA 02138, USA}
\email{melvis@cfa.harvard.edu}

\author[0000-0001-7134-9929]{Sunil P Khatri}
\affiliation{Department of Electrical and Computer Engineering, 3128 Texas A\&M University, College Station, TX 77843, USA}
\email{sunil@khatriweb.com}

\author[0000-0001-8325-7378]{Richard Linares}
\affiliation{Department of Aeronautics \& Astronautics, Massachusetts Institute of Technology, 77 Massachusetts Avenue, 33-207, Cambridge, MA 02139, USA}
\email{linaresr@mit.edu}

\author[0000-0002-7938-3432]{Manushaqe Muco}
\affiliation{Program in Media Arts and Sciences, MIT Media Lab, Massachussetts Institute of Technology, 77 Massachusetts Avenue, E14/E15, Cambridge, MA 02139, USA}
\affiliation{MIT Computer Science \& Artificial Intelligence Laboratory, Massachusetts Institute of Technology, 32 Vassar Street, Cambridge, MA 02139, USA} 
\email{manjola@mit.edu}

\author[0000-0002-6892-6948]{S.~Seager}
\affiliation{Department of Physics and Kavli Institute for Astrophysics and Space Research, Massachusetts Institute of Technology, Cambridge, MA 02139, USA}
\affiliation{Department of Earth, Atmospheric and Planetary Sciences, Massachusetts Institute of Technology, Cambridge, MA 02139, USA}
\affiliation{Department of Aeronautics and Astronautics, MIT, 77 Massachusetts Avenue, Cambridge, MA 02139, USA}
\email{professorsaraseager@gmail.com}

\author[0000-0001-6160-5888]{Jason T. Wright}
\affiliation{Department of Astronomy \& Astrophysics, The Pennsylvania State University, 251 Pollock Road, University
Park, PA 16802, USA}\affiliation{Penn State Extraterrestrial Intelligence Center, The Pennsylvania State University, 525 Davey Laboratory, 251 Pollock Road, Penn State, University Park, PA 16802, USA}
\affiliation{Center for Exoplanets and Habitable Worlds, The Pennsylvania State University, 525 Davey Laboratory, 251 Pollock Road, Penn State, University Park, PA 16802, USA}
\email{astrowright@gmail.com}

\begin{abstract}

The Dyson Minds 2025 Workshop, held at the Center for Brains, Minds \& Machines at MIT and organized by Penn State, MIT, and The Ultraintelligence Foundation, brought together researchers in astrophysics, engineering, artificial intelligence, computer science, and philosophy to examine ``Dyson Minds'' --- large-scale post-biological intelligences powered by energy harvested from supermassive black holes (SMBHs). Building on the ideas of \cite{dyson1960,dyson1966} and \cite{good1966}, participants explored the physical, engineering, behavioral, and observational consequences of civilizations embodied as machinery operating near the universe’s most powerful energy sources. The workshop aimed to develop new observational strategies capable of detecting signatures of such systems. Despite the highly cross-disciplinary scope, discussions centered on how a Dyson Mind might be constructed, how it might behave, and how those factors would shape strategies for the search for extraterrestrial intelligence. Key themes included the thermodynamic, mechanical, and stability limits of Dyson swarms; the trade-offs between power availability and communication latency in distributed minds; and how observability changes depending on whether Dyson Minds act as coherent entities or as loosely coordinated collectives. Across these topics, the consensus was that details of architecture and behavior strongly influence observational signatures. A major recommendation was to apply anomaly-detection methods to archival datasets, including those from WISE, JWST, and the Event Horizon Telescope, to identify unusual sources potentially overlooked by standard reduction pipelines. By integrating insights from multiple disciplines, the meeting advanced concrete, observation-focused strategies for future technosignature searches around SMBHs.

\end{abstract}


\section{Introduction and Background}

On June 3-4, 2025, an interdisciplinary assembly of experts convened at the Massachusetts Institute of Technology's (MIT) Center for Brains, Minds \& Machines to commence Dyson Minds 2025. This workshop was led by members of Penn State University, MIT, and The Ultraintelligence Foundation and consisted of $\sim25$ attendees from around the world. The goal of this workshop was to explore the possibility, nature, design, and detectability of large-scale computational intelligences dubbed ``Dyson Minds.'' These hypothetical entities are based on the foundational work of Freeman Dyson \citep{dyson1960,dyson1966} and Irving J. Good \citep{good1966}, representing advanced artificial intelligences operating at galactic or intergalactic scales. 

Although \cite{dyson1960} originally imagined a circumstellar ``biosphere'' that collects the entire energy output of a star, the idea was later generalized to mean any industrial use of large amounts of stellar energy \citep{dyson1966}. In some contexts, a civilization might progress to the point of exploiting the energy supply of its entire stellar system through the creation of a satellite swarm (colloquially known as a Dyson sphere) that collects starlight, processes it, and (because energy must be conserved) radiates away heat at mid-infrared wavelengths{, which, although not universally agreed upon, generally refers to wavelengths between $\sim3$ to $(25-40)$ microns (see, e.g., \citealt{stern2005} and \citealt{stern2012})}. Identifying waste heat from these objects has since become a major subfield in the search for extraterrestrial intelligence (SETI; e.g., \citealt{wright2014a, wright2020}). 

Many authors have imagined Dyson spheres as a tool for computation (see, e.g., \citealt{scharf2024}; \citealt{wright2024}) which extracts low-entropy energy from starlight, performs calculations, generates entropy from the resetting of system memory which it stores in the collected energy, and then expels this entropy as waste heat. {\cite{wright2024} argues that Dyson spheres operating at extremely cold temperatures are unlikely due to being inefficient to construct, which leaves the fiducial value for the temperature that a Dyson sphere operates at to be between $\sim 100$ to $\sim1000-3000$K (or $\sim1-30\mu $m, the latter of which is bounded by the melting point of your building material \citep{wright2020}. However, some authors have argued for the existence of very cold Dyson spheres (e.g., \citealt{Lacki2016}), which in turn might necessitate searches out to longer wavelengths.} {Still,} this framework has many advantages and may be universal: computation has proven on Earth to be a highly fungible and elastic good as ever-more-efficient generalized computers find new uses (e.g., \citealt{koomey2011,hilbert2011}). Space also provides an environment with enormous available volume for storage and infrastructure, essentially unlimited stellar energy, and no atmosphere or biosphere that would suffer adverse feedback from large-scale energy harvesting or waste-heat disposal \citep{dyson1960,Bostrom2003}.

A logical consequence of imagining Dyson spheres as computers stems from \cite{good1966} idea regarding the emergence of ultra-intelligence (now also called superintelligence). According to Good, because designing machines itself constitutes an intellectual activity, a sufficiently intelligent machine would be able to design even more intelligent machines. This capability would trigger a self-reinforcing process, rapidly amplifying intelligence to a point vastly beyond human levels, producing ultra-intelligent post-biological minds {(i.e., intelligences operating on non-biological, engineered substrates rather than organic cognition \citep{Dick2003}, although the exact definition of the term is under debate in both astrobiology and computer science \citealt{gershenson2025})}. 

As \cite{Dick2020} argues, a post-biological universe may be an inevitability of a developing culture due to the time-scales involved in cosmic evolution, the potential age of an {extraterrestrial intelligence (ETI)}, and the idea of an ``intelligence principle'' that posits the accumulation and propagation of knowledge as the main cultural impetuses. This presents an ``algorithm-gene-culture'' extension \citep{DickInPress} to the usual ``gene-culture'' coevolution of advanced civilizations \citep{Lumsden2005} where post-biological life becomes the driving force seeking and maintaining the information accrued by a culture. As \cite{Bostrom2024} argues, such post-biological life could arise as a system of accumulated agents, each with unique sub-goals and interests. It becomes clear that, in the context of Dyson Minds that are composed of individual swarm elements that may or may not act autonomously from each other, their nature becomes a fundamental aspect that is intertwined with their observability. 

The premise of this workshop was to explore the consequences of the hypothesis that such Dyson minds might be an inevitable consequence of biological life evolving into post-biological life, at least somewhere in the universe.  The goal of this workshop was to advance our understanding of the physical, engineering, computational, and behavioral dimensions of Dyson-scale intelligence, with the ultimate objective of setting observational constraints that can inform SETI investigations. 

Participants began their discussion by focusing on the observability of advanced intelligences that are harvesting energy from the most continuously luminous objects in the universe, i.e., the accretion disks around actively accreting supermassive black holes (SMBH; defined as black holes with masses between $10^6-10^9M_\odot${; see, e.g., \citealt{Shakura1973}; \citealt{Soltan1982}; \citealt{kormendy2013}; \citealt{heckman2014}; \citealt{netzer2015})}. Such systems offer unparalleled power densities ($L\sim10^{44-47} \; \rm{erg}\; s^{-1}$ for SMBH masses $\sim10^{6}-10^{9}M_\odot$) and long-term stability, making them an attractive energy source for any civilization seeking to maximize the free energy available for large-scale computation.  In doing so, it became obvious that the behavior, design, structure, and coherence of these minds directly affect their observability, and thus that SETI for post-biological intelligence must involve researchers and research across a broad range of fields.  

Dyson Minds raise \cite{vinge1993} ideas on \textit{singularity}, that is, a future time where an artificial mind develops to the point that it becomes impossible for humans to predict or understand its behavior. \cite{vinge1993} suggested that natural laws or evolutionary pressures might constrain such intelligences, but he emphasized that we cannot rely on our current intuitions about behavior, ethics, or societal organization to predict their choices. However, participants argued that, from a SETI perspective, the nature of minds and their behaviors can affect their observability, and that natural laws, such as evolution, might allow us to identify regions of a phase space of minds and behaviors that would be detectable using specific search strategies. 

To emphasize this, consider two scenarios of Dyson Minds in orbit around a {SMBH that is actively accreting at or near the Eddington accretion rate ($\sim10^{45} \rm{erg \: s^{-1}}$, or around $10^9$ times more luminous than the black hole at the center of our galaxy \citealt{sgrAStar}).} One Dyson Mind might try to conceal its presence in the galactic center to avoid attracting competition. However, such a civilization is still subject to the laws of thermodynamics and must choose what to do with its waste heat \citep{wright2020}, potentially beaming the energy along astrophysical jets outside of the plane of the galaxy, where it is less likely to directly interact with an observer. Such a mind would thus sacrifice efficiency to reduce detectability, obfuscating its presence with natural phenomena, so searchers for these objects will need to be focused on delineating abnormal modulations from natural stochastic processes.

On the other hand, consider a Dyson Mind that seeks to boisterously establish itself at a prime energy source to actively intimidate other civilizations from attempting to gain comparable footholds. The SMBH of a galaxy would be a very tantalizing location for such a mind due to its very high energy density and perhaps its centralized location. A Dyson Mind might then adopt a ``Spread and Suppress'' strategy,  aiming to be the first to arrive at the SMBH so it can establish a strong foothold. The suppression of other civilizations, either as it developed its way towards the central engine or after it has established itself there and seeks to monopolize the energy output, might also leave detectable signatures, such as a trail of megastructures left behind around energy sources under its control. With regards to observability, such a mind might bombastically broadcast its presence in hopes of dissuading competition from approaching its energy source. Traditional waste heat, radio, and laser SETI may thus be effective in detecting such minds.

 

This workshop was thus a cross-disciplinary endeavor, bringing together expertise that spanned astrophysics, engineering, artificial intelligence, philosophy, fundamental physics, ecology, and computer science with the goal of understanding how the behavior and engineering constraints of Dyson Minds might affect SETI. To facilitate such a broad discussion, the workshop was structured over two days, with the first day involving introductory presentations and a series of interdisciplinary discussions, referred to as ``verticals'' and ``horizontals,'' that were focused on specific thematic and cross-cutting issues, respectively. That is, the discussions were grouped along the framework termed ``PEABO'':

\begin{enumerate}
    \item P (Physics): Astrophysical constraints, thermodynamics, and energetics of Dyson swarms.

    \item E (Engineering): Practical considerations in computing and infrastructure at astronomical scales.

    \item A (AI): Computational architectures and algorithms suitable for distributed intelligence.

    \item B (Behavior): Societal, evolutionary, and competitive dynamics of Dyson-scale intelligences.

    \item O (Observability): Identifying and interpreting potential technosignatures for observational searches.
\end{enumerate}

\noindent The verticals were mixed-discipline groups where all fields (physics, engineering, AI, behavior, observability) were represented. Verticals were meant to explore the space of Dyson Mind possibilities broadly, with cross-disciplinary input, and each vertical session used different participant mixes to encourage idea diversity. Horizontals, on the other hand, were single-focus groupings where participants were grouped by related expertise. In this workshop, two horizontal tracks ran in parallel:
\begin{enumerate}
    \item Physics and Engineering (PE) - A focus on the design and physical constraints underlying a Dyson Mind.
    \item Brains and minds (AB) - A focus on possible structures and behaviors of very large coherent computational minds. 
\end{enumerate}

\noindent Day 1 of the workshop had 3 sessions: a vertical, a horizontal, and a vertical. The horizontals were intended to dig deeper into discipline-specific issues that were raised in the verticals, so that specialists could clarify technical points for the broader group before the next mixed vertical. Day 2 centered on bringing ideas from all of the verticals and horizontals together in order to synthesize discussions on the observability of Dyson Minds.

The rest of this manuscript is structured as follows. In \S\ref{sec:corethemes} we discuss the central themes of this workshop, including the physical constraints (\S\ref{sec:physconstraints}), computational constraints (\S\ref{sec:latencyandalgorithms}), and ``taxonomy'' of Dyson Minds (\S\ref{sec:nature}). How these considerations affect observational strategies is presented in \S\ref{sec:observability}. Lastly, we conclude with a discussion and summary of the workshop in \S\ref{sec:conclusion}.

\section{Core Scientific Themes and Outcomes} \label{sec:corethemes}
\subsection{Physical and Thermodynamic Constraints} \label{sec:physconstraints}

No matter how advanced, a Dyson Mind will still face physical and thermodynamic limitations on its computational systems. First, Dyson spheres must be swarms, not shells, to avoid mechanical and gravitational instabilities \citep{wright2020}. Such systems require active station keeping and debris control to prevent Kessler cascades \citep{lacki2025}, implying the need for continuous collision avoidance and propulsion. The latter might not necessarily come from fuel, as, for small individual components, radiation pressure on the panels and exhaust from the radiators are sufficient for station keeping. Without such mechanisms, \cite{lacki2025} has shown that such swarms would succumb to instabilities on timescales as short as a few orbital periods.

Where to place a Dyson sphere around a SMBH also does not have a trivial solution. Computers in space only have access to radiative cooling, which places tight constraints on the accessible power density at a given component temperature. \cite{wright2020} and \cite{wright2024} argue that, while computing is more efficient at lower temperatures, more compact and therefore hotter Dyson spheres achieve a higher computational rate per unit mass. Thus, a Dyson Mind may wish to put its components as close to the central engine as possible, with minimum distances set by the need to keep individual swarm components from melting. 

Around a SMBH, the temperature of the radiation field is primarily set by the accretion disk that extends to $\sim1$au from the central engine \citep{Shakura1973, Soltan1982, Netzer2013}. For an extremely active galactic nucleus (AGN), the luminosity of these accretion disks can be equal to, or even be in excess of, the Eddington Luminosity \citep{King2003, Marconi2004}. For a SMBH of mass $10^6M_\odot$ accreting at the Eddington luminosity, a radiator with a surface area of $10^4\;\rm{cm}^2$ would need to be located $\sim1$pc away to achieve compute temperatures of $\sim300$K. Meanwhile, observations of accretion flows around Sgr A* (i.e., the SMBH at the center of the Milky Way) indicate that its energy output is $\sim10^{-7}-10^{-8}$ the Eddington luminosity \citep{Genzel2010, Yuan2014}. In this case, a Dyson Mind can be as close as $\sim10^2$au. {Operating around a lower-luminosity SMBH therefore substantially reduces the required collecting area needed to intercept a fixed fraction of the available power, easing material and station-keeping demands. An important caveat to this, however, is that a more luminous SMBH produces a higher radiation field density such that, at a fixed orbital radius and collecting area, a Dyson Mind might more easily meet its energy needs with a smaller collecting area around a hotter black hole. Still, the weaker radiation field also mitigates UV and X-ray damage, radiation pressure effects, and wind-driven ablation, improving long-term structural survivability. From an observational perspective, it would also be harder to detect a Dyson Mind stationed around a less luminous SMBH, potentially aiding a concealment-oriented Dyson Mind.}

The outer limits of the swarm configuration are set by the inner radius of the dust-obscuring region that surrounds most SMBH \citep[see][for a review]{hickox2018}. Within this region, the density of dust grains would significantly obscure a large portion of the collectible flux and increase the likelihood of a collisional cascade. 
These winds, along with intense UV and X-ray fluxes, would rapidly degrade structural materials and photovoltaics, especially those operating near the sublimation limit \citep{Elvis2022}. Over time, such exposure would erode panels and radiators, reducing reflectivity, absorptivity, or emissivity, and thus both power collection efficiency and radiative cooling performance. The timescales for uncontrolled degradation are uncertain. Short ablation timescales (i.e., thousands of years) could be feasible in high-flux regions, leading to complete ablation of panels in $<1$ million years. A long-lived Dyson Mind structure would thus need an ability to maintain eroded surfaces. One strategy could involve sacrificial shielding layers that absorb the brunt of the wind particle flux, at the cost of a cut to radiative flux onto its panels, but even these shields would need to be maintained, likely in a self-regenerating fashion.

Even in the absence of a Dyson Mind, the vicinity around a SMBH is a crowded, dynamically complex environment (see, e.g., \citealt{eckart1996}; \citealt{genzel1997}; \citealt{ghez1998}). Tens of thousands of stars, including massive progenitors and compact remnants, may orbit within the sphere of influence of the black hole. Many of these remnants are themselves products of supernova explosions, which not only inject intense radiation into the local environment but can also perturb individual Dyson Mind swarm elements, potentially leading to a Kessler cascade (see, e.g., \citealt{lacki2025}). Depending on their proximity, a supernova itself may ablate surfaces, disrupt thermal management systems, and destroy poorly shielded collectors or radiators. 

A sufficiently advanced Dyson Mind may use the stellar population around a SMBH to its advantage. Careful manipulations of the orbit of a star in orbit around a SMBH (see, e.g., \citealt{Dyson2023}) could send the star into the SMBH. This so-called ``stoking of the fire'' would manipulate the accretion flow onto the SMBH, allowing the Dyson Mind to garden the energy output of the compact object. A Dyson Mind that has had time to prepare may also choose to clear this stellar neighborhood before taking up residence around the SMBH. 

\subsection{Communication Latency and Computational Paradigms} \label{sec:latencyandalgorithms}


Communication latency will be a crucial challenge that any Dyson Mind-like entity will need to overcome. Around a sun-like star, a Dyson Mind would face latencies of $\sim10$ minutes, but around a SMBH, nodes could face latencies on scales ranging from days to multiple years. In fact, latency may matter more than total available power for determining whether a Dyson Mind functions as a coherent single intelligence or as a looser federation of sub-minds. Communication latency in Dyson-scale systems is not merely an engineering inconvenience but is fundamentally constrained by the laws of physics.

{The finite speed of light sets an irreducible lower bound on signaling time, while thermodynamic and energetic limits constrain both computation and communication rates. Foundational work on the physical nature of information \citep{Landauer1961} and on entropy bounds in finite systems \citep{Bekenstein1981} establishes that information processing is inseparable from energy flow and physical scale. Similarly, ultimate limits to computation and communication imposed by energy density and relativistic signaling have been explored by \cite{Bremermann1962} and \cite{lloyd2000}. In distributed systems embedded in deep gravitational potentials, additional effects such as relativistic time dilation and geometric light-travel delays further complicate synchronization \citep{Novikov1973}. These considerations make it clear that large-scale coherence in a Dyson Mind must contend not only with engineering trade-offs but with fundamental bounds set by relativity, thermodynamics, and information theory.}

A singularly coherent Dyson Mind will thus need to balance a trade-off between compact, lower‑latency compute and extended, high‑power swarms (greater latency). This trade underlies expectations about behavior and signatures over time. The former could range from a singular compute node to a single ring of nodes in low orbits, sacrificing covering fraction, and thus available power, in favor of maintaining a centralized core. In the latter case, as the swarm develops to incorporate more rings of compute elements, the Dyson Mind will need to transfer information from node-to-node over vast distances, potentially on scales of multiple lightyears.

This problem is not without potential solutions. For instance, one could allow physical nodes to orbit freely and instead keep any piece of information in a fixed spatial position. This could be done by transmitting the data between passing nodes in sync with orbital motion, or more simply, by transmitting the data to the swarm element behind it in its ring. This could keep logical locality stable despite the rapidly changing distributed system; however, it would require a significant portion of power to be put into maintaining these transmissions and performing error correction when necessary. 

Of course, for truly massive datasets, physical transport beats bandwidth in some regimes. It can be faster and more energy-efficient to load data onto a physical medium and send it on a trajectory to the destination than to transmit it via EM signals, especially when the signal bandwidth is fundamentally limited by thermal noise, diffraction, or available power. This would be analogous to the ``sneakernets'' on Earth, whereby shipping storage drives physically, rather than sending the data electronically, can often be more efficient. Orbital dynamics could be exploited to make this practical: nodes already move at high velocities around the central energy source, and carefully timed flybys or dedicated data couriers could transfer payloads without excessive propulsion requirements. While this method does achieve a higher throughput of information, it does incur higher absolute latencies than electromagnetic links and requires flybys at high relative velocity, which may present traffic-management dangers. 

A mixture of both models in the form of a multi-tiered communication system could also be possible. In this framework, urgent, low-volume messages travel via electromagnetic signals while bulk, non-time-critical datasets are moved physically between regions of the swarm. Such an architecture naturally lends itself to hierarchical or hub–dominated structures—similar to other efficient communication and transport systems that exhibit scale-free or power-law connectivity as a consequence of optimization and self-organization \citep{barabasi1999, carlson1999, barthelemy2011}. Large swaths of data can thus be sent on ``couriers'' on eccentric or highly inclined orbits to distant parts of the swarm, 
while time-critical messages can be sent through electromagnetic channels. {For example, for a $10^7M_\odot$ black hole surrounded by a Dyson Mind that is 1 parsec in radius, the light-crossing time from the center to the shell is $\sim 3.3$ years, whereas the Keplerian orbital period at this radius is $\sim 3 \times 10^4$ years.}

{To drive this point home further, consider the fact that the maximum rate of information transfer, $C$, through a noisy electromagnetic channel is bounded by the Shannon–Hartley theorem \citep{Shannon1949}, which gives $C=B\log_2(1+S/N)$, where B is the bandwidth and  S/N the signal-to-noise ratio. Even for an optimistic 1 GHz channel with moderate signal-to-noise, capacities are of order a few Gb$\: s^{-1}$. Transmitting even a modest 10 TB hard drive would require around a full day of constant transmission at these rates. We note that the transmission of electromagnetic signals depends heavily on the total power that a Dyson Mind allocates to the task as well as its chosen wavelength. Both of these facets would affect detectability and how we might detect a transmission that is beamed along our line of sight (e.g., \citealt{Fan2025}).}

{By contrast, information storage in matter is constrained by thermodynamic and gravitational bounds that are extraordinarily large. In the extreme case, the Bekenstein bound \citep{Bekenstein1981} and related limits on computation and information density \citep{lloyd2000} show that macroscopic matter can encode vastly more information per unit mass than can be economically embedded in an electromagnetic channel. In fact, if the courier has a sufficiently dense data volume, $I$, the dynamical time becomes comparable to the time it would take to beam an equivalent amount of data. This occurs when}

\begin{equation}
    I = 2 \pi C \sqrt{\frac{R^3}{GM}} \:,
\end{equation}

\noindent {where R is the radius at which the swarm elements are orbiting, G is the gravitational constant, and M is the mass of the black hole. For our above example, this occurs when $I\sim1$ zettabyte, well within the Bekenstein bound, highlighting the fact that physical couriers might be a very ubiquitous mode of data transfer. We note that if the Dyson Mind is able to form a physical connection between its swarm elements, then it would be able to achieve even faster data transfer rates. However, the strains that would be induced on the connectors (e.g., \citealt{wright2020}), as well as the fact that these connections will need to survive any station keeping maneuvers, means such connections may only be possible in certain swarm configurations.}


A Dyson Mind might operate computational nodes in different modes, with some calculations performed by massively parallel networks, slow serial computation done ``on the spot'' of energy generation, and other high serial rate calculations done less efficiently at concentrated nodes to which energy is beamed and where aggressive cooling mechanisms rapidly redistribute the waste heat to radiators.

Regardless of the hardware architecture, it is clear that parallel algorithms with sparse global coordination are the most natural computational model for Dyson-scale systems. From an engineering perspective, this constraint effectively forces a hierarchical architecture in which large-scale minds may need multiple ``rates of cognition'' to exist, with fast local processes operating independently and slower global reasoning integrating over long timescales. Practically, this could be achieved by incorporating fast controllers near the radiators and collectors, a handful of mid‑scale coordinators per ring, and a global synchronizer if coherence is important to the Mind. Over time, the slower layers could shape the goals and incentives of the faster ones, while the faster layers provide the raw computation required to achieve these global strategies.

To put it simply, a Dyson Mind is likely never to have tight synchronization across the whole system, so it should instead be designed such that it does not need it. Algorithmically, this could be achieved with a ``map locally, reduce rarely'' paradigm whereby the Dyson Mind focuses on doing as much work as possible within the immediate neighborhood of each node, globally synchronizing only when results require it. Evolutionary or genetic algorithms \citep{holland1992, back1993} may have intrinsic incentives for Dyson Minds, although no consensus was reached regarding how important these algorithms are for observability. 

Such algorithms are inherently well-suited to environments with high communication latency because they operate through many independent trials, local selection, and occasional information exchange. In the context of Dyson Minds, the periodic recombination of solutions could occur when orbital geometry brings certain nodes into closer communication windows, allowing information exchange without requiring the entire swarm to halt for a global sync. An advantage of this method is its high fault tolerance. That is, if a node fails for whatever reason, the distributed nature of the algorithm ensures that overall progress continues. Moreover, diversity of solutions is preserved naturally because distant regions evolve under different conditions and selective pressures, allowing a Dyson Mind to adapt to local environmental conditions or computational subtasks while still contributing to the global project.

Another viable strategy would exploit mixture-of-experts architectures, where many independent sub-models work in parallel and only occasionally exchange summaries or parameters. In these models, many specialized ``experts'' (e.g., neural networks, symbolic solvers, or other modules) run independently, each handling a subset of inputs or problem domains. A gating mechanism determines which expert handles which problem instance. In a Dyson Mind, this gating could occur locally due to physical spatial constraints. In this scenario, the results are routed when relevant experts need them, bypassing the need to broadcast all ideas across a network. Experts operate largely autonomously, and experts may go long periods without external input, only periodically contributing results or receiving updates during scheduled synchronizations. As such, these algorithms are not ideal if a single coherent mind is the desired goal.

Latency considerations thus suggest a strong incentive toward local specialization with sparse global integration. Evolutionary algorithms achieve this through decentralized search and selection, while mixture-of-expert algorithms achieve this through specialized functional partitioning. In practice, a combination of paradigms (e.g., an evolutionary search within and among experts, with occasional global policy updates) could be particularly powerful for a Dyson Mind constrained by high latencies.

\subsection{Nature, Origin, and Evolution of Intelligence} \label{sec:nature}

The taxonomy of Dyson Minds has major implications for their observability. Below, we discuss three potential branches in this taxonomic tree. That is, whether the Dyson Mind is a coherent or incoherent entity and whether it was an engineered or emergent Mind.

The former branch asks whether the Dyson Mind manifests as either a unified intelligence or as a collection of loosely organized subsystems. So-called coherent Dyson Minds are ones in which the computational and infrastructural components remain bound together as a singular entity. Achieving such coherence would require extraordinary continuity of purpose across tens of thousands of years, since the construction of a large-scale swarm is necessarily incremental. Should such a mind also be engineered (see below), the stability of its progenitor civilization is also a key factor in its survival. In such coherent systems, there may in fact be no limit to the maximum size of a brain, provided the architecture can adapt to ever-growing latency constraints. This view suggests that coherent minds could, in principle, scale to enormous sizes, functioning as a single cognitive entity so long as their internal organization remains synchronized with the physics of communication.

By contrast, an incoherent Dyson Mind was envisioned less as a single unified intelligence and more as an ecosystem of interacting agents. Such a system could consist of multiple layers of organization, with entities of different sizes and levels of influence coexisting within the same swarm. Such a system could consist of many computational clusters, cultural lineages, or even competing entities sharing only minimal communication protocols. Instead of functioning as one mind, it would resemble a distributed network in which diverse subsystems interact, sometimes cooperating and sometimes competing, creating a dynamic landscape of overlapping cognitive structures. The result would be an ensemble of minds, variable in size and loosely linked, each with its own local coherence. These Dyson Minds may be more reminiscent of an ecosystem of semi-autonomous subsystems rather than a single conscious whole, whereby different subminds may pursue distinct strategies while still contributing to a broader, loosely federated structure. Such a collection would nonetheless have emergent behaviors, similar to how an ant colony or humanity is composed of many discrete minds that exhibit collective intelligence, memory, and purpose.

In summary, coherent Dyson Minds can be thought of as incrementally built integrated systems that might, under certain architectures, achieve measurable consciousness. Incoherent Dyson Minds were depicted as distributed ecosystems of subminds that were interconnected, albeit lacking deep integration. In either case, the motivation of the Dyson Mind, or the society that constructed it, is also critical to consider.

Along with being sources of vast amounts of energy, BHs are also a unique astrophysical laboratories. An exploration-based entity may very well find itself around a central SMBH in hopes of running fundamental physics experiments that require exorbitant amounts of energy (see, e.g., \citealt{lacki2015}). Such experiments would produce vast amounts of nonthermal radiation, especially in the form of YeV neutrinos \citep{lacki2015} or non-stochastic accretion flows \citep{Vidal2011}, that would not otherwise be produced by natural accretion onto the black hole {\citep{Soltan1982}}. The centers of galaxies are thus important targets of interest for high-energy SETI. However, some Dyson Minds may eschew such experiments, instead focusing on efficient energy acquisition, so such searches for YeV radiation must necessarily be complementary to other search methods. {Primordial black holes \citep{Zeldovich1966} have also been thought of as promising targets for Dysonian structures \citep{Baghram2025}, where they would leave noticeable signals on observed microlensing light curves \citep{Baghram2026}.}

Both scenarios provide very different implications for how such entities behave and how they might be observed, but they are not the only deciding factors at play. For instance, consider whether or not the Mind is an engineered structure, which may or may not inherit the motivations of those who created it, or an emergent mind whose motivations are decoupled from any outside civilization, will both lead to different technosignatures. For instance, the Dyson Minds in the former example are the result of deliberate large-scale construction projects initiated by an advanced civilization. Their architecture would likely reflect intentional abstraction layers and modular subsystems, much like human-built computational infrastructures, allowing for maintainability, upgrades, and clear separation of functions. These systems might be optimized for efficiency, resilience, and redundancy, with carefully tuned trade-offs between energy harvesting, latency, and computational throughput. From an observational perspective, engineered Dyson Minds could produce technosignatures that are highly structured and patterned, such as regular power modulations or geometric patterns in transit light curves (see, e.g., \citealt{arnold2005}), making them potentially easier to distinguish from natural astrophysical processes. 

On the other hand, emergent Dyson Minds are the product of a more natural evolutionary process. In this scenario, the mind may not even be a single coherent entity and instead takes the shape of fragmented entities that are decoupled from one another. However, it should be noted that an emergent Dyson Mind could still be coherent, especially since any swarm will likely need some form of coordinated space situational awareness and station keeping. Emergent entities could begin as autonomous machines or distributed AI nodes. Interactions between such structures would be shaped by local interactions, perhaps even competition, among subsystems, leading to architectures that may be inefficient, redundant, and heterogeneous. Observationally, emergent Dyson Minds could be much harder to detect as there might not be any of the regularities or peculiar geometries that are present in the designed case. Evolutionary time-scales also play an important factor, since, if the transition times for the emergence are short, there will only be a few galaxies for which we could observe this transition. In the worst case, the stochasticity of interwoven minds may make their technosignatures indistinguishable from natural astrophysical phenomena. However, their incoherence may also be a boon to observability, as the actions of any one subsystem are unpredictable and may choose to broadcast its location, potentially even against the desire of the Dyson Mind as a whole \citep{wright2014a}. This could lead to an ecology of swarms where individual or groups of entities are in competition with other swarm components.

The answers to either of these questions (i.e., coherent vs. incoherent or engineered vs. emergent) are neither dependent on one another nor mutually exclusive. An engineered Dyson Mind could be coherent or incoherent, and an otherwise coherent Dyson Mind could partition part of itself to be its own incoherent entity. The latter of which represents an engineered subsystem of a potentially emergent global Dyson Mind. Nevertheless, the answers to these questions fundamentally dictate the nature of a Dyson Mind, but different combinations of these answers may lead to differing observing strategies, thus warranting careful consideration of each topic.

\section{Observability and Open Research Directions} \label{sec:observability}

A central goal of the workshop was to identify observational pathways by which Dyson Minds might be detected or ruled out. The discussions repeatedly emphasized that observability is not only a matter of physical principle but also of architecture (\S\ref{sec:physconstraints}), communication strategy (\S\ref{sec:latencyandalgorithms}), and cognitive organization (\S\ref{sec:nature}). A few broad categories of technosignatures emerged.

The first referred to waste heat and infrared emission. Dyson-scale computation must reradiate nearly all of the energy it absorbs, and thus mid-infrared thermal emission remains the most robust technosignature. The characteristic temperature and spectrum are determined by the placement of collectors, the radiative cooling efficiency, and the nature of the dust in the surrounding environment, particularly for Minds operating near SMBHs. Compact configurations favoring low latency (see \S\ref{sec:latencyandalgorithms}) radiate from hotter, smaller areas, while power-maximizing extended swarms glow more brightly in the mid-infrared. Observationally, this suggests searching for anomalous excesses, unusual color–color outliers, or smooth blackbody-like components inconsistent with dust emission. A caveat to this is the fact that some Dyson Minds may instead choose to store energy locally (e.g., into the SMBH spin), redirect it to off-site facilities where it might be radiated at lower temperatures (see \S\ref{sec:physconstraints}). In such cases, the apparent infrared output would underestimate the true energy budget of the system, complicating searches based solely on highly localized waste heat.

Beyond broadband thermal emission, Dyson Minds may introduce structured anomalies into their host environments. {These could include departures from standard AGN spectral templates, such as, modification of jet energetics or morphology or unusual elemental abundances detected around the AGN due to an increase in amount of iron and silicate (i.e., metals typically found in computers) to the system.} Cross-band correlated variability may be particularly diagnostic: engineered systems could display fixed time lags, quasi-periodic modulation, or synchronized behavior across emission regions that differ from the stochastic variability of natural accretion flows. Individual components of a Dyson Mind may also scatter light off their surfaces into our line of sight. Unlike the polarization that arises from synchrotron emission or scattering off of dust, engineered surfaces could display wavelength-dependent polarization features tied to materials or geometries, producing distinctive polarization curves. Moreover, swarms orbiting in front of the inner accretion disk could produce transient eclipses in the X-ray or UV flux. These dips would differ from natural AGN absorption variability by their sharp ingress and egress timescales, repeatability, or coherence across multiple wavelengths \citep{Imara2018, DiStefano2021}. In either case, the key signature would be structured, periodic, or otherwise non-stochastic.

Dyson Minds may not even confine themselves to local computation around the SMBH, instead opting to transmit information to off-site computation facilities (see \S\ref{sec:latencyandalgorithms}). In such cases, tightly focused or lasers could act as technosignatures. Directionality and beam steering would produce anisotropic, patterned signals. However, a Dyson Mind attempting to hide its presence may cloak these signals, disguising itself as astrophysical phenomena.

At larger scales, Dyson Minds could reshape their surroundings in ways that are detectable even in the absence of direct technosignatures. Abandoned swarms, for example, may decay into dust within many times the orbital timescale, leaving little to no long-term trace (\citealt{lacki2025}). Conversely, active ``Spread and Suppress'' strategies might generate population-level anomalies, such as regularized patterns of infrared excess across multiple stars, suppression of competition in occupied regions. A Dyson Mind that is actively ``stoking the fire'' by actively accreting nearby stars onto the SMBH may eventually leave appreciable gaps in the center of the galaxy. So-called ``deforestation'' scenarios could be detectable by identifying galaxies whose surface brightness profiles deviate significantly from, say, a S\'ersic profile \citep{sersic}.

The jets of some AGN have also long been thought of as potential energy sources (see, e.g., \citealt{hsiao2021}). After all, these jets can carry away significant fractions of the radiation of the accretion disk in the form of collimated, relativistic plasma streams. However, even low-power jets reach sizes of $\sim 1$kpc \citep{Cavagnolo2010}, potentially making their sizes impractical for Dyson Minds seeking a compact energy source. Still, proposed ideas include placing solar sail-like collectors downstream of the jet, keeping it fixed in place by balancing the inward gravitational force with the ram pressure of the jet. However, it is unclear whether any material can survive the ablation induced by a relativistic plasma stream. Certainly, any Dyson Mind swarm component collecting radiation off the accretion disk will need to avoid any jets, and the Dyson Mind itself may attempt to garden its SMBH so as not to produce a jet. The mechanisms behind jet formation are not well characterized, but the luminosity of a jet is known to be highly correlated with the accretion flow (see, e.g., \citealt{ghisellini2014}). So-called ``knots'' of high-density plasma that have been observed along many AGN jets (see, e.g., \citealt{lister2016}; \citealt{jorstad2017}) are thought to be the downstream ejecta from tidal disruption that occurred near the accretion disk. As such, in situations where a Dyson Mind is actively gardening the accretion flow onto a SMBH, we may be able to observe the effects of this in the form of ``exhaust'' being shot out along the jet. Reanalyzing Very Long Baseline Array data of the jet kinematics of AGN could reveal such gardening in the form of knots that appear to have been emitted periodically.  

Dyson Mind observability may be a long-tail phenomenon, meaning Dyson Minds that are most detectable will not necessarily be the most common, but rather those whose architectures produce conspicuous signatures, whose behaviors favor broadcasting over concealment, or whose scale is so extreme that they cannot be hidden. Conversely, Dyson Minds that prioritize efficiency, or those that actively cloak their emissions, may remain invisible, detectable only through indirect means. In either case, the search for Dyson Minds must be framed as the search for astrophysical outliers: rare, distinctive deviations from overwhelmingly natural backgrounds. However, it has also been hypothesized that the number of Kardeshev Type III civilizations is either extremely common or nearly nonexistent \citep{Papagiannis1980, Kipping2025}, in which case the search for ``outliers'' could also be the search for the few galaxies that do not contain a Dyson Mind. In this case, observational strategies that compare populations of galaxies (e.g., groups of galaxies) against one another may be useful in detecting outliers.

The systematic reprocessing of archival astronomical data through the lens of anomaly detection may be a lucrative way forward for the field. Infrared surveys such as the Wide-field Infrared Space Explorer (WISE; \citealt{Wright2010}) already provide wide baselines for color-space outlier searches (see, e.g., \citealt{suazo2022}), but there are numerous datasets that have never been examined through a SETI perspective. In the context of looking for Dyson Minds, the {Event Horizon Telescope (EHT; e.g., \citealt{EHTI}; \citealt{EHTII}}) has produced unprecedented horizon-scale images of SMBH environments. These data could be reanalyzed to search for Dyson Minds around M87 and Sgr A*. In particular, EHT data could be used to search for swarm components passing in front of the compact object or anomalous polarization measurements that differ from the stochastic synchrotron background. 

{A critical next step is forward modeling of AGN spectral energy distributions (SED) that incorporate partial interception of accretion disk luminosity by a Dyson Mind that is radiating at $T\sim100-3000$K. For back of the envelope example, consider a $10^7M_\odot$ black hole radiating at 10\% of the Eddington luminosity ($L_D\sim 10^{44}\rm{erg \: s^{-1}}$. A Dyson Mind with a covering fraction of $f\sim10\%$ would thus reradiate $10^{43}\rm{erg\:s^{-1}}$ in the form of infrared radiation (assuming that the Dyson Mind does not beam or store any of the energy). This would give an effective emitting radius of}

\begin{equation}
    R_{\rm eff} = \sqrt\frac{f L_D}{4 \pi \sigma T^4} \sim 1.3 \;\rm{pc},
\end{equation}

\noindent {where $\sigma$ is the Stefan-Boltzmann constant. At 300K, such a component would be comparable in luminosity to the mid-infrared output of a typical dust-obscured region (e.g., \citealt{honig2017}; \citealt{honig2017}), potentially reshaping the $5-20\mu$m SED in amplitude, slope, and variability behavior. Even modest interception fractions therefore produce significant mid-infrared signatures. Complete forward modeling of AGN SEDs would thus let us predict these behaviors and allow us to attain complete posterior probability distributions on swarm temperatures and covering fractions, and it could also inform the ranges in mid-infrared color-space that an AGN surrounded by a Dyson Mind might occupy (e.g., \citealt{stern2005}; \citealt{stern2012}).}

{For JWST, the most relevant instrument is the Mid-Infrared Instrument (MIRI; \citealt{miri}). Broad-band imagaing with MIRI would be highly sensitive to excess thermal emission in the $5-28\mu$m range and could identify AGN that are color outliers relative to torus models. MIRI Medium Resolution Spectroscopy (MRS) provides additional leverage as the mid-infrared emission from a Dyson Mind would appear as a smooth blackbody-like continuum, potentially diluting or suppressing the typical silicate and hydrocarbon features that are seen in AGN spectra \citep{draine2007}. Spectroscopy is therefore essential for distinguishing engineered thermal emission from clumpy dust reprocessing.}

{Hotter Dyson Mind configurations ($T\gtrsim800-1000$K would shift peak emission toward $2-4\mu$m, making them accessible to the Near Infrared Camera (NIRCam; \citealt{rieke2005}) instrument on JWST and to wide-field near-infrared surveys such as Euclid \citep{euclid} and the Nancy Grace Roman Space Telescope \citep{roman}. Wide near-infrared surveys could therefore identify AGN with anomalously hot thermal components or unusual near-infrared excesses that are inconsistent with standard torus templates. In practice, an effective strategy would combine wide-field near-infrared color selection with targeted MIRI spectroscopy to fully constrain whether candidate objects exhibit smooth thermal continua, anomalous variability, or other departures from standard AGN SED models.}

It should be noted that, while SMBHs were chosen as the primary location for a Dyson Mind for the purposes of this workshop, they were not the only location considered. In particular, accretion onto stellar mass black holes (e.g., X-ray binaries) can reach luminosities as large as $\sim10^{38}-10^{41}\rm{erg}\;s^{-1}$ \citep{Ryu2016}. Although much less than the energy output of accretion onto a SMBH $\gtrsim10^{44}\rm{erg}\;s^{-1}$, it could be a lucrative stepping stone for a civilization or Dyson Mind looking to upgrade from the energy output of its host star. 

Lastly, there could be a hierarchy of targets to search for. A Dyson Mind might progress from orbiting a young star to expanding to nearby stars or stellar mass black holes before surrounding itself around the SMBH at the center of its galaxy. \cite{wright2021} modeled that a civilization in a Milky Way-like galaxy is highly dependent on the peculiar motions of stars it expands to. The differential motion of stars throughout the galaxy produces a spiral settlement wave moving inward, implying that a Dyson Mind might take a similar path on its route towards the center of a galaxy. Analysis of waste heat searches of resolved galaxies could find similar spiral patterns in the form of technosignatures left behind by the settlement wave.

\section{Discussion and Summary} \label{sec:conclusion}

The Dyson Minds 2025 Workshop explored the concept of large-scale computational intelligences operating at astronomical scales, building upon Dyson’s and Good’s foundational ideas by imagining so-called Dyson Minds as physically constrained, engineered systems whose behaviors leave potentially detectable signatures. Doing so was a monumental effort that required a cross-disciplinary effort from the fields of physics, astronomy, machine learning, artificial intelligence, computer engineering, computer science, and ecology. Determining the parameter space of observable features of a Dyson Mind was the cornerstone of all discussions, and, in the end, this workshop was successful in balancing pure speculation against calculations that would allow for the upper detectability limits to be set on particular behaviors and architectures. 

A fundamental assumption was that Dyson Minds must operate within the bounds of known physics. Such an entity must necessarily be swarms rather than shells \citep{wright2020}, requiring station keeping, collision avoidance, and an ability to survive the intense winds and radiation that can be found around a SMBH. We posit that a Dyson Mind will seek the largest available energy gradient in the universe --- that of the SMBH at the center of its galaxy.


A recurring theme was the tension trade-off between latency and raw power collection. From a computational perspective, minds around a SMBH might not operate as a tightly synchronized entity, instead adopting multiscale architectures consisting of fast, local modules handling reflexive control and maintenance, and slower, distributed layers integrating knowledge over long timescales. Evolutionary algorithms, mixture-of-experts systems, and other locality-friendly approaches were seen as natural fits for this environment, as they all incentivize integration between local specialization and global decision-making.

Two critical splits in the taxonomic tree of Dyson Minds were heavily discussed (i.e., whether the mind was engineered or an emergent phenomenon and whether it is a coherent or incoherent mind). Although neither dependent on nor mutually exclusive, the answer to either of these questions can alter observational strategies. For example,  engineered Dyson Minds may reflect the logic and modularity of their creators, producing structured, patterned technosignatures. Meanwhile, emergent Dyson Minds may be inefficient, redundant, and indistinguishable from natural phenomena, but this could also lead to the scenario of an emergent mind that is interested in broadcasting its existence out to the galaxy \citep{wright2014a}. In that vein, the motivation of the Dyson Mind also affects observability. 

Observability discussions highlighted a range of potential technosignatures. Waste heat remains the most robust, but this method has caveats due to the fact that some Dyson Minds may attempt inefficient methods of energy storage or redirection.
Spectral and variability anomalies, including polarization signatures and X-ray eclipses, may prove to be equally important, as any engineered surface transiting the central engine will both scatter and polarize light and produce sharp X-ray eclipses. At the largest scales, entire environments could be reshaped or ``gardened'' to meet the needs of the Dyson Mind. For instance, the stars in the vicinity of the SMBH could have their orbits purposefully perturbed in a way that triggers a tidal disruption event. Like throwing fuel onto a fire, such a practice could be used to increase the accretion rate, and thus total luminosity, of the SMBH accretion disk. If done properly, this method would also impart angular momentum onto the SMBH that can later be extracted {(e.g., \citealt{Penrose1969}). Still, before any direct searches for Dyson Minds can be conducted, forward models for their effects on SMBH and AGN observables must be developed and tested.} 

A clear near-term recommendation was to reprocess existing astronomical datasets. Facilities such as WISE, the, EHT, {and the James Webb Space Telescope (JWST; \citealt{JWST})} already produce data that can be tested for anomalies in spectral energy distributions, variability patterns, and polarization. Indeed, archival searches have already been applied to search for stellar Dyson spheres \citep{suazo2022, Suazo2024} and extragalactic technosignature searches \citep{wright2014a,wright2014b,Griffith2015}, and in general, can lead to anomalous discoveries that would otherwise have been missed \citep{Hashimoto2019, Lacki2021, Etsebeth2024}. In a general scientific context, \cite{Humphreys1966} argues that the study of anomalies is necessary for major scientific discoveries --- a theme \cite{Cleland2019} argues will be required to develop a universal theory of life \cite{Cleland2022}. Applying anomaly detection frameworks to the search for Dyson Minds could reveal signatures hidden in plain sight. 

Another direction that was discussed and is being contemplated for implementation is the development of a "Dyson Minds Simulator" (DMS). The DMS would draw from the expertise of the attendees from all the areas. It would simulate Dyson Minds of different scales and sizes, while
simultaneously modeling multiple constraints such as computation architectures and hardware, algorithms, power delivery and power dissipation, and communication latencies. This would allow for the exploration of several what-if scenarios and provide insight into the
stability of such scenarios. A possible beneficial side effect of the DMS development effort would be to come up with new computing paradigms that might be relevant for the Internet of Things (IoT) computational framework for present-day terrestrial computing.

In conclusion, Dyson Minds may be inevitable consequences of evolution, but their detectability depends on how they negotiate the trade-offs between power, latency, and efficiency, the role of evolution and competition, and other factors to be explored. Placing observability constraints on Dyson Minds is thus not a problem that a single field can answer, but is instead a complex issue that requires the synthesis of physics, engineering, computation, and behavior. The workshop demonstrated that by treating Dyson Minds as constrained but diverse physical systems, SETI can move beyond speculation toward concrete, testable strategies for identifying signs of extreme intelligence operating around the most powerful energy sources in the present Universe.

\section{Acknowledgments} 

{We would like to thank the anonymous reviewer for their insightful comments. }We would also like to acknowledge all attendees of the Dyson Minds 2025 Workshop, especially: Mark Beal (MarbleSemi), Gaby Contardo (University of Nova Gorica, Slovenia), Sheperd Doeleman (Harvard), Cecilia Garraffo (Harvard), Paolo Gaudenzi (University of Rome), Paul Horowitz (Harvard), Brian Lacki (Breakthrough Listen, University of Oxford), Richard Linares (MIT), Lakshminarayanan Mahadevan (Harvard), Tom Malone (MIT), Tommy Poggio (MIT), Anders Sandberg (Institute of Future Studies), Susan Schneider (Florida Atlantic University), Steinn Sigurdsson (Penn State), Gerry Sussman (MIT), Jesse Thaler (MIT).

The workshop was supported by The Ultraintelligence Foundation, Penn State, and MIT. O.C. acknowledges support from The Ultraintelligence Foundation and the Penn State Extraterrestrial Intelligence Center.

\newpage 

\bibliography{PASPsample701}{}
\bibliographystyle{aasjournalv7}

\end{document}